\documentclass[utf8,pra,twocolumn,amsmath,amssymb,floatfix,reprint,footinbib,superscriptaddress,longbibliography,showkeys]{revtex4-2}
\pdfoutput=1
\usepackage{CJK}
\usepackage{dcolumn}
\usepackage{graphicx}
\usepackage{mathrsfs}
\usepackage{mdwlist}
\usepackage{subfigure}
\usepackage{booktabs}
\usepackage{amsmath}
\usepackage{extarrows}
\usepackage{dsfont}
\usepackage{amstext}
\usepackage{amssymb}
\usepackage{amsbsy}
\usepackage{bbm}
\usepackage{bm}
\usepackage{amsthm}
\usepackage{graphicx}
\usepackage{textcomp}
\usepackage{multirow}
\usepackage{color}
\usepackage{diagbox}
\usepackage{braket}
\usepackage{appendix}

\usepackage{xcolor} 
\usepackage{soul}
\soulregister\cite7 % 针对\cite命令 
\soulregister\ref7 % 针对\ref命令 
\soulregister\eqref7

\usepackage[colorlinks,citecolor=blue]{hyperref}
\definecolor{Dgreen}{RGB}{0, 100, 0}
\usepackage{url}
\usepackage[colorlinks]{hyperref}
\hypersetup{%
	plainpages=true,
	breaklinks=true,  %not default in dvips mode, so we must specify
	hypertexnames=false,  %not ideal, but needed when pagenums duplicate (`i' vs. `1')
	pageanchor=true,
	colorlinks=true,
	linkcolor={blue},
	citecolor={blue},
	urlcolor={blue},
	anchorcolor={black}
}

\begin{document}
\title{Detecting a single atom in a cavity using the $\chi^{(2)}$ nonlinear medium}

\author{Dong-Liang Chen}\thanks{Both authors contributed equally to this work.}
\affiliation{Fujian Key Laboratory of Quantum Information and Quantum Optics (Fuzhou University), Fuzhou 350108, China}
\affiliation{Department of Physics, Fuzhou University, Fuzhou 350108, China}

\author{Ye-Hong Chen}\thanks{Both authors contributed equally to this work.}
\affiliation{Theoretical Quantum Physics Laboratory, RIKEN Cluster for Pioneering Research, Wako-shi, Saitama 351-0198, Japan}

\author{Yang Liu}
\affiliation{Fujian Key Laboratory of Quantum Information and Quantum Optics (Fuzhou University), Fuzhou 350108, China}
\affiliation{Department of Physics, Fuzhou University, Fuzhou 350108, China}

\author{Zhi-Cheng Shi}
\affiliation{Fujian Key Laboratory of Quantum Information and Quantum Optics (Fuzhou University), Fuzhou 350108, China}
\affiliation{Department of Physics, Fuzhou University, Fuzhou 350108, China}

\author{Jie Song}
\affiliation{Department of Physics, Harbin Institute of Technology, Harbin 150001, China}

\author{Yan Xia}\thanks{xia-208@163.com}
\affiliation{Fujian Key Laboratory of Quantum Information and Quantum Optics (Fuzhou University), Fuzhou 350108, China}
\affiliation{Department of Physics, Fuzhou University, Fuzhou 350108, China}

\begin{abstract}
We propose a protocol for detecting a single atom in a cavity with the help of the $\chi^{(2)}$ nonlinear medium. When the $\chi^{(2)}$ nonlinear medium is driven by an external laser field, the cavity mode will be squeezed, and thus one can obtain an exponentially enhanced light-matter coupling.~Such a strong coupling between the atom and the cavity field can significantly change the output photon flux, the quantum fluctuations, the quantum statistical property, and the photon number distributions of the cavity field.~This provides practical strategies to determine the presence or absence of an atom in a cavity.~The proposed protocol exhibits some advantages, such as controllable squeezing strength and exponential increase of atom-cavity coupling strength, which make the experimental phenomenon more obvious. We hope that this protocol can supplement the existing intracavity single-atom detection protocols and provide a promise for quantum sensing in different quantum systems.
\end{abstract}

\maketitle
\section{INTRODUCTION}\label{INTRODUCTION}
Cavity quantum electrodynamics (cavity QED) system~\cite{dutra2005cavity,haroche2006exploring} is one of the most promising platforms for realizing quantum information processing.~It is mainly used to study the dynamical behavior of particles (e.g.,~atoms and ions) confined to a high-finesse cavity interacting with photons~\cite{dutra2005cavity,haroche2006exploring}.~Moreover, atoms have the advantages of high coherence and long lifetime.~Therefore, atom-cavity-coupled systems become an excellent choice for experiments in quantum information and quantum computation.~Studying the interaction between various light fields and atoms in a cavity is an important research field of quantum optics~\cite{weiner2008light,scully1997quantum}.~Up to now, many models have been proposed to study the light-matter coupling, for examples, the Jaynes-Cummings (JC) model~\cite{1443594,doi:10.1080/09500349314551321}, the Tavis-Cummings (TC) model~\cite{PhysRev.170.379}.~Based on these models~\cite{1443594,doi:10.1080/09500349314551321,PhysRev.170.379}, the interaction of different types of atoms with various light fields~\cite{PhysRevA.35.154,GOU1990218,6106827,obada1987time,PhysRevA.101.053826,PhysRevA.102.013714} has been extensively studied and many interesting quantum optical phenomena have been probed, such as vacuum Rabi splitting~\cite{yoshie2004vacuum}, squeezing phenomena of optical fields~\cite{doi:10.1080/09500348714550721,PhysRevA.37.3175,PhysRevA.81.015804}, single photon blockade~\cite{birnbaum2005photon}, and so on.~Furthermore, many schemes for quantum information and quantum computation using atom-cavity-coupled systems have been proposed, for examples, realizations of quantum gates~\cite{PhysRevA.56.3187,PhysRevLett.74.4087}, generations of entangled states~\cite{PhysRevLett.85.2392,PhysRevA.53.2857,PhysRevA.98.042310,PhysRevA.91.012325,PhysRevA.96.033803,PhysRevA.95.022317,PhysRevA.89.052313,PhysRevA.102.053118}, operations of a quantum phase gate~\cite{PhysRevLett.83.5166}, and so on~\cite{PhysRevLett.83.4204,PhysRevA.67.042311,Shi:18,PhysRevA.102.022617,Yu-chi-ZHANG:190,Shuai-Liu:21502,PhysRevA.101.012329}.

In the above schemes~\cite{PhysRevA.56.3187,PhysRevLett.74.4087,PhysRevLett.85.2392,PhysRevA.53.2857,PhysRevA.98.042310,PhysRevA.91.012325,PhysRevA.96.033803,PhysRevA.95.022317,PhysRevA.89.052313,PhysRevA.102.053118,PhysRevLett.83.5166,PhysRevLett.83.4204,PhysRevA.67.042311,Shi:18,PhysRevA.102.022617,PhysRevA.102.022617,Yu-chi-ZHANG:190,Shuai-Liu:21502,PhysRevA.101.012329}, one premise of realizing quantum information and quantum computation using atom-cavity-coupled systems is to trap atoms in a cavity.~Therefore, the ability to nondestructively detect the presence of atoms in a cavity is very important for quantum information processing.~This problem has attracted considerable attentions.~Up to now, there have been proposed several single-atom detection schemes~\cite{Kuhr278,schlosser2001sub,lev2004feasibility,Bao:19,PhysRevB.90.134515,goldwin2011fast,Haase:06,ott2016single,PhysRevLett.98.233601,PhysRevA.67.043806,PhysRevLett.97.023002}.~For instance, detecting atoms in dipole traps using fluorescence detections~\cite{Kuhr278,schlosser2001sub}, detecting a single atom using photonic bandgap cavities~\cite{lev2004feasibility}, sensing single atoms in a cavity using broadband squeezed light~\cite{Bao:19}, qubit measurement by interferometry with parametric amplifiers~\cite{PhysRevB.90.134515}.~In addition, the enhanced coupling between atoms and photons inside a high-finesse optical cavity provides a novel basis for optical measurements that continuously monitor atomic degrees of freedom.~The real-time detection of a single cold atom falling through a high precision optical cavity has been realized~\cite{Mabuchi:96,PhysRevLett.80.4157}.~Observing a single atom in a blue-detuned intracavity dipole trap, the blue ``funnels'' demonstrated in this scheme could efficiently  guide an atom to regions of strong atom-cavity coupling,  thereby enhancing the detection efficiency in the experiments of single-atom detection~\cite{PhysRevLett.99.013002}.~However, these schemes have some drawbacks, for example, at present, the direct coupling of broadband squeezed vacuum field and cavity is hard to realize experimentally~\cite{Bao:19}, and the squeezed strength is difficult to control, which may hinder the implementations of the experiment.

To date, the nonlinear media has attracted the attention of researchers because of their special functions, and has been widely used in nonlinear optics~\cite{PhysRevLett.12.504,PhysRev.182.482,PhysRevLett.14.973,1070043,https://doi.org/10.1002/andp.202000002,qin2020strong,AsghariNejad:17,PhysRevLett.12.592,boyd2020nonlinear}.~For example, it can be used to generate the stimulated Raman scattering~\cite{PhysRevLett.12.504,PhysRev.182.482}, optical parametric oscillations~\cite{PhysRevLett.14.973}, and amplifications~\cite{1070043,https://doi.org/10.1002/andp.202000002,qin2020strong,Yexiong-Zeng:12503}.~Moreover, the nonlinear medium also can be used to observe a variety of nonlinear optical effects, such as the optical Kerr effect~\cite{AsghariNejad:17}, the stimulated Brillouin scattering~\cite{PhysRevLett.12.592}, and so on.~In particular, the $\chi^{(2)}$ nonlinear optical medium~\cite{boyd2020nonlinear} is commonly used for optical frequency doubling, mixing, and optical parametric oscillation effects.~Recently, $\chi^{(2)}$ nonlinear mediums have been widely used in quantum optics due to their ability to produce parametric amplification effect~\cite{PhysRevLett.120.093601,PhysRevLett.120.093602,PhysRevLett.126.023602,burd2021quantum,PhysRevA.100.012339,PhysRevLett.114.093602,lemonde2016enhanced,PhysRevA.100.062501}.~Therefore, in order to avoid the drawbacks of single-atom detection in a cavity mentioned above~\cite{Bao:19}, using a $\chi^{(2)}$ nonlinear medium to detect the single atom in a cavity may be a good idea.

In this paper, we propose a scheme for detecting a single atom in a cavity using the $\chi^{(2)}$ nonlinear medium.~The $\chi^{(2)}$ nonlinear medium is driven by an external laser field to produce a squeezed-light field.~This can exponentially enhance the coupling between an atom and a squeezed-light field, which changes some physical properties of the system, such as the output photon flux~\cite{PhysRevA.100.062501}, quantum fluctuations, quantum statistical properties, and photon number distribution.~Therefore, one can determine whether a single atom is present in the cavity by observing changes in the physical properties of the system.~Note that our scheme uses parametric amplification to enhance the coupling strength, whereas Refs.~\cite{Mabuchi:96,PhysRevLett.80.4157,PhysRevLett.99.013002} vary coupling strengths by adjusting detunings.~Moreover, the squeezing strength and squeezed-cavity-mode frequency can be adjusted by tuning the amplitude of a driving field or the detuning between cavity field frequency and additional driving field frequency. Therefore, the scheme may be relatively easy to achieve in the experiment. In a word, the scheme is feasible to detect a single atom in a cavity using the $\chi^{(2)}$ nonlinear medium, and it has the advantages of obvious experimental phenomena, low implementation difficulties, and controllable experimental parameters.

This paper is structured as follows.~In Sec.~$\rm\ref{MODEL}$, we describe the physical model and give the Hamiltonian of the system.~In
Sec.~$\rm\ref{METHOD}$, we propose a scheme for detecting a single atom in a cavity with the help of the $\chi^{(2)}$ nonlinear medium. Finally, the conclusion is given in Sec.~$\rm\ref{CONCLUSION}$.

\section{PHYSICAL MODEL}\label{MODEL}
\begin{figure}[!htbp]\centering
	\scalebox{0.45}{\includegraphics[width=1\textwidth]{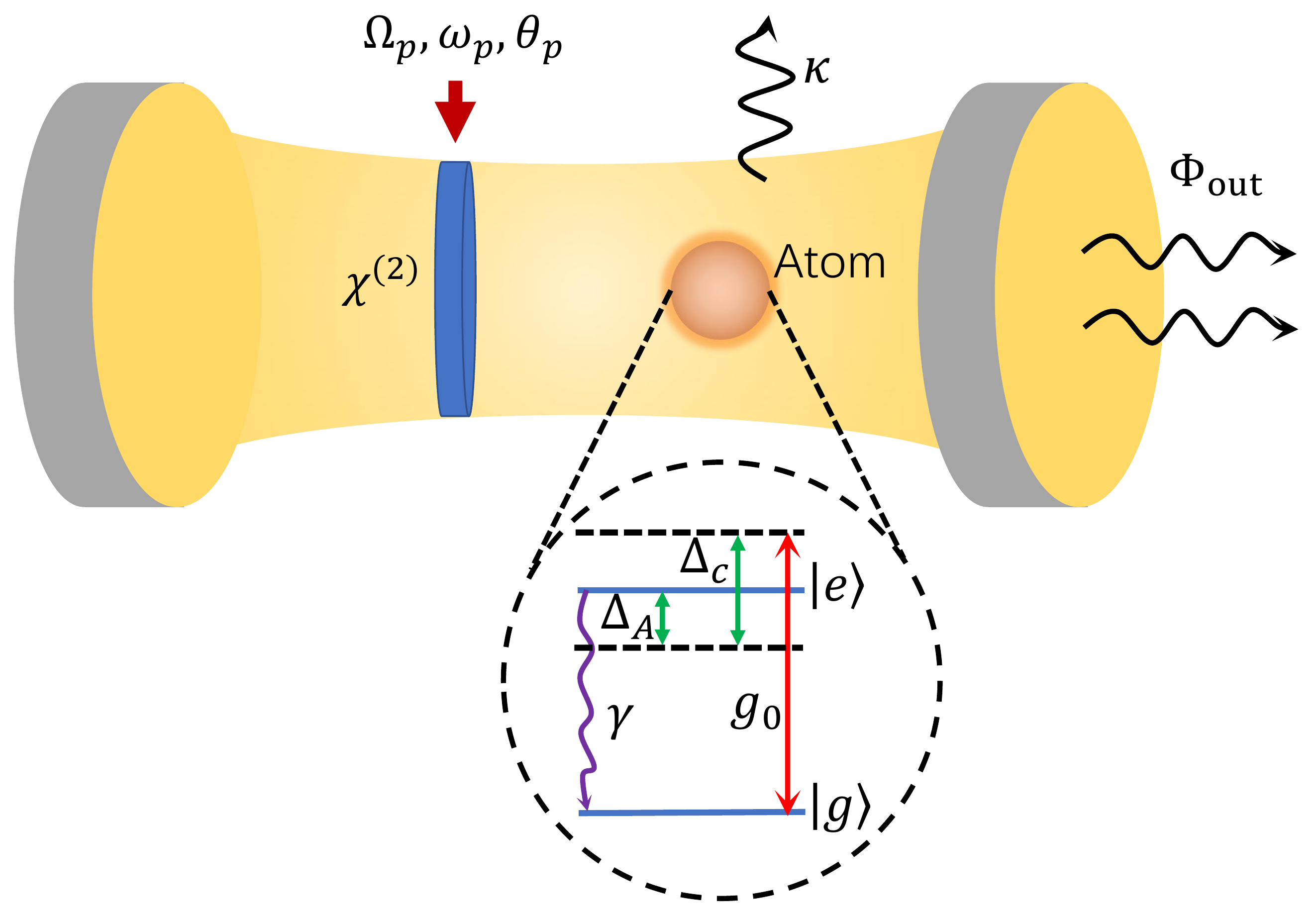}} \caption{Schematic illustration of a cavity QED system
		containing a single-mode cavity, a two-level atom, and a $\chi^{\left( 2\right) } $ nonlinear medium.
		The two-level atom, with a ground state $\ket{g}$ and an excited state $\ket{e}$, is trapped in a single-mode cavity and coupled to the cavity with an atom-cavity coupling strength $g_0$.
		The $\chi^{\left( 2\right) } $ nonlinear medium is strongly pumped at amplitude $\Omega_p$, frequency $\omega_p$, and phase $\theta_p$.
		Here,~$\gamma$ and $\kappa$ are the decay rates of the atom and cavity, respectively.}
	\label{Fig1}
\end{figure}

As shown in Fig.$~\ref{Fig1}$, we consider a cavity QED system containing a two-level atom, a $\chi^{\left( 2\right) } $ nonlinear medium, and a single-mode cavity. The atom, with a ground state $\ket{g}$ and an excited state $\ket{e}$, is confined in a single-mode cavity with frequency $\omega_c$. In this cavity QED system, the $\chi^{\left( 2\right) }$ nonlinear medium is driven by an additional driving field with frequency $\omega_p$, amplitude $\Omega_p$, and phase $\theta_p$, resulting in a two-photon effect, which is used to squeeze the cavity mode. The detunings are $\Delta_c=\omega_c-\omega_p/2$ and $\Delta_{A}=\omega_A-\omega_p/2$, where $\omega_{A}=\omega_e-\omega_g$ is the atomic transition frequency.

To be specific, we consider the Hamiltonian of system in a proper observation frame (hereafter, we set $\hbar=1$)
\begin{equation} \label{eqn1}
	\begin{aligned}
		H&=\Delta_{A}\sigma_{ee}+H_{\rm AC}+H_{\rm NL},\\
		H_{\rm AC}&=g_0\left(a\sigma_{eg}+ a^{\dagger}\sigma_{ge}\right),\\
		H_{\rm NL}&=\Delta_{c}a^{\dagger}a+ \frac{1} {2}\Omega_p\left[ {\rm exp}(i\theta_p)a^2+\rm H.c.\right],
	\end{aligned}
\end{equation}
where, $H_{\rm NL}$ is the nonlinear Hamiltonian for degenerate parametric amplification, $H_{\rm AC}$ is the Hamiltonian of the atom-cavity coupling, $g_0$ is the coupling strength between the atom and cavity, $a$ and $a^\dagger$ denote the annihilation and creation operators of cavity field, respectively. The two-level atom is described by Pauli operator $\sigma_{ee}=\ket{e}\bra{e}$ and transition operators $\sigma_{eg}=\sigma_{ge}^{\dagger}=\ket{e}\bra{g}$, where $\ket{g}$ and $\ket{e}$ are the ground state and excite state, respectively. In order to simplify calculating, we assume that $\theta_p=0$ in the following discussions.
According to $H_{\rm NL}$ in Eq.~(\ref{eqn1}), when the $\chi^{\left( 2\right) }$ nonlinear medium is driven, the photons in the cavity will be produced or annihilated in pairs.

The evolution of the system can be determined by a master equation in the Lindblad form
\begin{equation} \label{eqn2}
	\begin{aligned}
		\dot{\rho}(t)=&i\left[\rho(t),H\right]\\
		&-\frac{1}{2}\gamma\left[ \sigma_{ee}\rho(t)-2\sigma_{ge}\rho(t)\sigma_{eg}+\rho(t)\sigma_{ee}\right]\\
		&-\frac{1}{2}\kappa\left[a^{\dagger}a\rho(t)-2a\rho(t)a^{\dagger}+\rho(t)a^{\dagger}a\right],
	\end{aligned}
\end{equation}
where, $\rho(t)$ is the density operator, $\kappa$ is the dissipation rate of the cavity, and $\gamma$ is the dissipation rate of the two-level atom from $\ket{e}$ to $\ket{g}$. When no atom is trapped in the cavity, the dissipation of the cavity and two-photon  effect leads to the energy level transitions.~On the contrary, when the atom is trapped in the cavity, atomic spontaneous emission and atom-cavity coupling will generate additional transition paths.

\begin{figure}[!htbp]\centering
	\scalebox{0.45}{\includegraphics[width=1\textwidth]{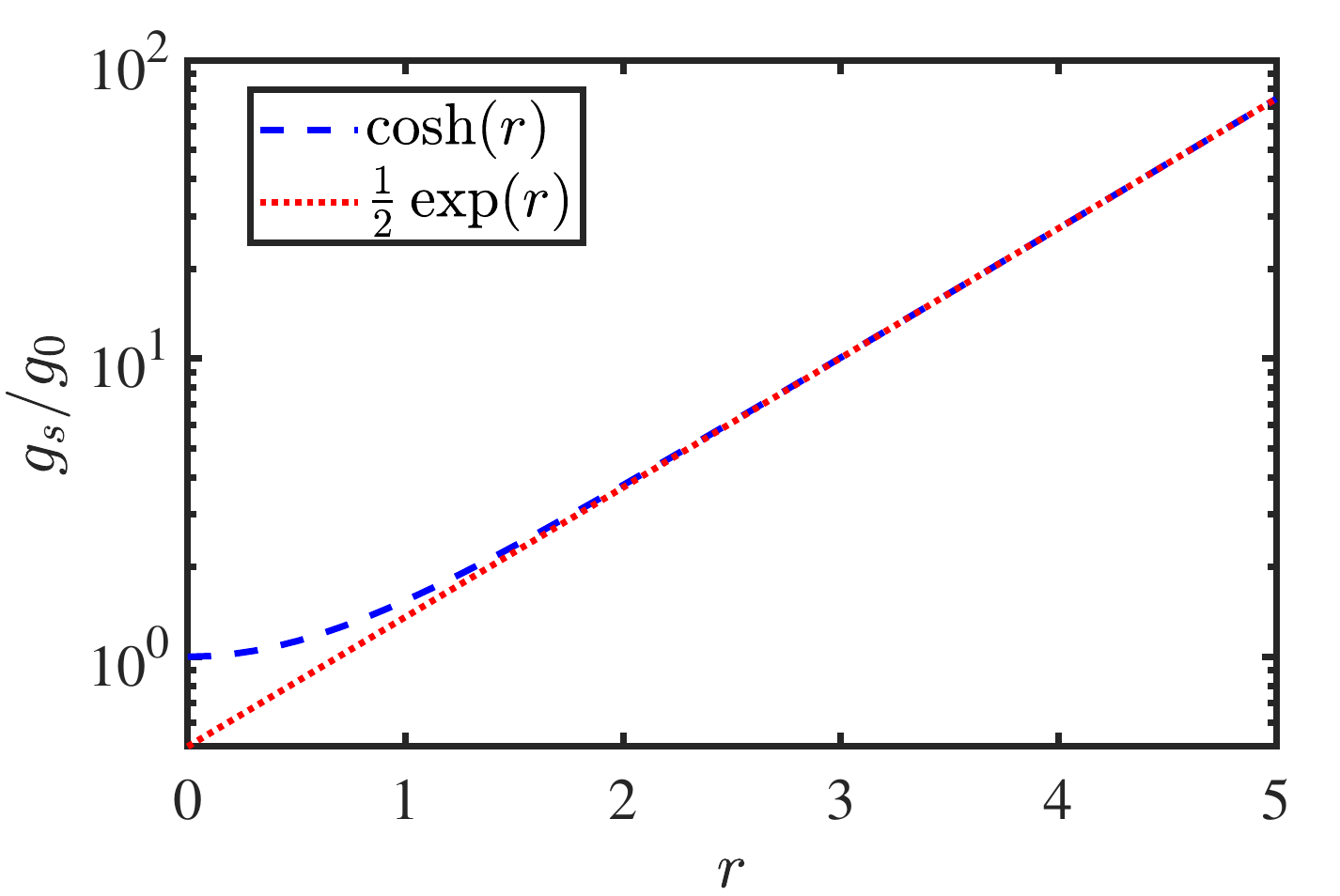}} \caption{The enhance ratio $g_s/g_0$ of atomic-cavity coupling strength versus the squeezing parameter $r$.~The blue dashed curve and the red dotted curve denote ${g_s}/{g_0}=\cosh(r)$ and ${g_s}/{g_0}\sim\frac{1}{2}\exp(r)$ of the enhancement ratio $g_s/g_0$, respectively.}
	\label{Fig2}
\end{figure}

When driving the $\chi^{(2)}$ nonlinear medium by the additional driving field $\Omega_p$, the bare cavity mode $a$ can be transformed to a squeezed mode $a_s$ with the squeezing parameter(see Appendix~$\rm\ref{A}$ for details)
\begin{eqnarray}\label{e3}
	r=\frac{1}{4}\ln\left(\frac{\Delta_c+\Omega_p}{\Delta_c-\Omega_p} \right).
\end{eqnarray}
As a result, the nonlinear Hamiltonian $H_{\rm NL}$ in Eq.~($\rm\ref{eqn1}$)
is diagonalized as
\begin{eqnarray}\label{e4}
	H_{\rm NLS}=\omega_sa_s^{\dagger}a_s,
\end{eqnarray}
by the Bogoliubov squeezing transformation~\cite{scully1997quantum}
\begin{eqnarray}\label{e5}
	a_s=\cosh(r)a+\sinh(r)a^\dagger,
\end{eqnarray}
where, $\omega_s=\sqrt{\Delta_c^2-\Omega_p^2}$ is a controllable squeezed-cavity-mode frequency, which depends on the detuning $\Delta_c=\omega_c-\omega_p/2$ and the amplitude $\Omega_p$.~After substituting Eq.~(\ref{e5}) into Eq.~(\ref{eqn1}), the Hamiltonian $H_{\rm ACS}$ of the interaction between the atom and the squeezed cavity mode becomes
\begin{eqnarray}\label{e6}
	H_{\rm ACS}=\left( g_sa_s-g_s^{\prime}a_s^{\dagger}\right)\sigma_{eg}+\left(g_sa_s^{\dagger}-g_s^{\prime*}a_s\right)\sigma_{ge},
\end{eqnarray}
where $g_s=g_0\cosh(r)$ and $g_s^{\prime}=g_0\sinh(r)$. When $|g_s^{\prime}|/(\omega_s+\Delta_{A})\ll1$ and $\omega_s=\Delta_{A}$, the counter-rotating terms of Eq.~($\ref{e6}$) can be neglected by the rotating-wave approximation. As a consequence, $H_{\rm ACS}$ can be rewritten as
\begin{eqnarray}\label{e7}
	H_{\rm ACS}^{\prime}= g_s\left(a_s\sigma_{eg}+a_s^{\dagger}\sigma_{ge}\right).
\end{eqnarray}
As shown in Fig.$~\ref{Fig2}$, when $r\geq1$, the blue dashed curve and the red dotted curve are basically identical, that is, the coupling strength $g_s$ between the atom and the squeezed cavity mode increases exponentially with increasing the squeezing parameter $r$,
\begin{eqnarray}\label{e8}
	\frac{g_s}{g_0}=\cosh(r)\sim\frac{1}{2}\exp(r).
\end{eqnarray}
This exponential enhancement is explained by the fact that the parameter drive changes the eigenstates of the cavity Hamiltonian and the photon is squeezed into a squeezed photon with amplified fluctuations, resulting in greater interaction with the atom~\cite{PhysRevLett.120.093602,PhysRevA.101.053826}. Furthermore, this enhancement is not equivalent to simply injecting compressed light into the cavity without changing the Hamiltonian of the system.
Similar approaches have been used to improve the light-matter interactions and cooperativity of cavity QED systems~\cite{PhysRevLett.120.093601,PhysRevLett.120.093602,PhysRevLett.126.023602,burd2021quantum,PhysRevA.100.012339} or optomechanical systems~\cite{PhysRevLett.114.093602,lemonde2016enhanced,PhysRevA.100.062501}.

Based on the above analysis, the Hamiltonian $H$ of the system in Eq.~(\ref{eqn1}) in the squeezed frame can be expressed as
\begin{eqnarray}\label{e9}
	H_s=\Delta_{A}\sigma_{ee}+\omega_sa_s^{\dagger}a_s+g_s\left(a_s\sigma_{eg}+a_s^{\dagger}\sigma_{ge}\right).
\end{eqnarray}
By contrast, when no atoms are present in the cavity, the Hamiltonian of the system in the squeezed frame is $H_{\rm NLS}$.

In the squeezed frame, the master equation that determines the evolution of the system is expressed as
\begin{equation} \label{eqn11}
	\begin{aligned}
		\dot{\rho_s}(t)=&i\left[\rho_s(t),H_s\right]\\
		&-\frac{\gamma}{2}\left[ \sigma_{ee}\rho_s(t)-2\sigma_{ge}\rho_s(t)\sigma_{eg}+\rho_s(t)\sigma_{ee}\right]\\
		&-\frac{\kappa}{2}(N_s+1)\left[a_s^{\dagger}a_s\rho_s(t)-2a_s\rho_s(t)a_s^{\dagger}+\rho_s(t)a_s^{\dagger}a_s\right]\\
		&-\frac{\kappa}{2}N_s\left[a_sa_s^{\dagger}\rho_s(t)-2a_s^{\dagger}\rho_s(t)a_s+\rho_s(t)a_sa_s^{\dagger}\right]\\
		&+\frac{\kappa}{2}M_s\left[ a_s^{2}\rho_s(t)-2a_s\rho_s(t)a_s+\rho_s(t)a_s^{2}\right]\\
		&+\frac{\kappa}{2}M_s^*\left[ a_s^{\dagger 2}\rho_s(t)-2a_s^{\dagger}\rho_s(t)a_s^{\dagger}+\rho_s(t)a_s^{\dagger 2}\right],
	\end{aligned}
\end{equation}
where,
\begin{equation} \label{eqn12}
	\begin{aligned}
		N_s&=\sinh^2(r),\\
		M_s&=\cosh(r)\sinh(r),
	\end{aligned}
\end{equation}
describe thermal noise and two-photon correlations introduced into the cavity by squeezing, respectively.

\begin{figure}[!htbp]\centering
	\scalebox{0.45}{\includegraphics[width=1\textwidth]{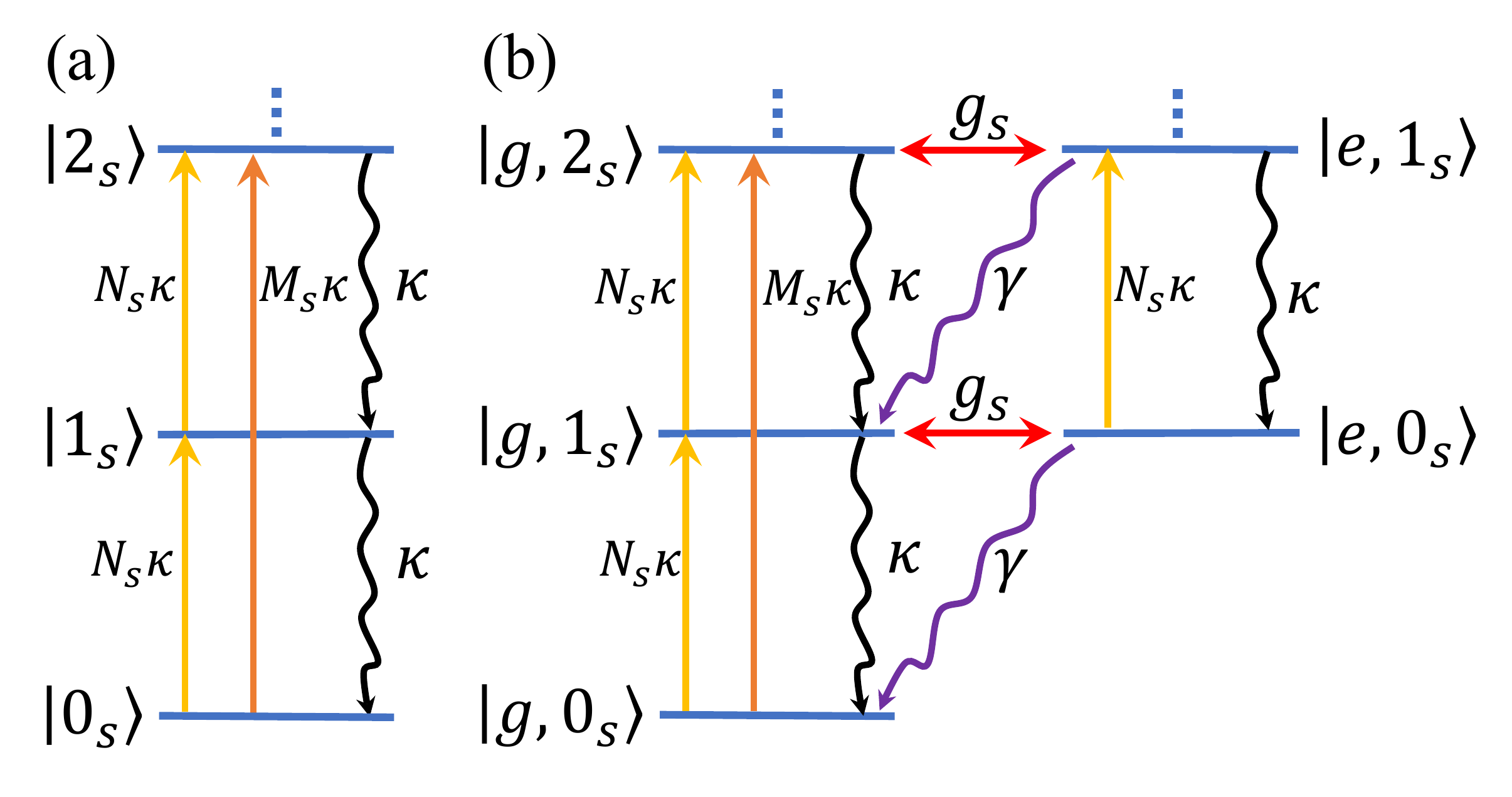}} \caption{The energy level structures and the transition pathways for (a) the empty cavity and (b) the single-atom-cavity QED system, within the squeezed frame.~Here, $\ket{n_s} (n=1, 2, \cdots N)$ represents that there are $n$ squeezed photons in the  squeezed cavity field and the symbol ``~$\vdots$~'' represents the higher energy levels.
}\label{Fig3}
\end{figure}

For the sake of clearness, in the following sections, we will refer to the cases of the presence and the absence of an atom in the cavity as the single-atom-cavity QED system and the empty cavity, respectively. In the squeezed frame, the energy level structures and the transition paths of the empty cavity and the single-atom-cavity QED system~\cite{Li:18}, are shown in Fig.$~\ref{Fig3}$.~That is, when no atom is trapped in the cavity, the energy level transitions are generated by the dissipation of the cavity and the thermal noise and two-photon correlation, see Fig.$~\rm\ref{Fig3}$(a). Here, the thermal noise and two-photon correlation are caused by compression. On the contrary, when the atom is trapped in the cavity, the atomic spontaneous emission and the atom-cavity coupling will generate additional transition paths, see Fig.$~\rm\ref{Fig3}$(b).

According to the above analysis, the next section will present in details how to detect a single atom in a cavity using the $\chi^{(2)}$ nonlinear medium.

\section{DETECTING A SINGLE ATOM IN THE CAVITY}\label{METHOD}
In this section, we discuss how to detect a single atom in a cavity with two-photon driving and the $\chi^{(2)}$ nonlinear medium.~We first study the dynamical evolution of the single-atom-cavity QED system with the squeezed frame.~Assuming that the cavity mode is initially in the squeezed-vacuum state $\ket{0_s}$ and the atom is initially in the ground state $\ket{g}$.~Here, $\ket{0_s}$ is a superposition of only even-photon number states~\cite{scully1997quantum,drummond2013quantum},
\begin{equation} \label{e13}
	\begin{aligned}
		\ket{0_s}=\frac{1}{\sqrt{\cosh r}}\sum_{n=0}^{\infty}(-1)^n\frac{\sqrt{(2n)!}}{2^nn!}(\tanh r)^{n}\ket{2n}.
	\end{aligned}
\end{equation}

\begin{figure}[!htbp]\centering
	\subfigure{
		\label{Fig4:a}  \includegraphics[width=0.23\textwidth]{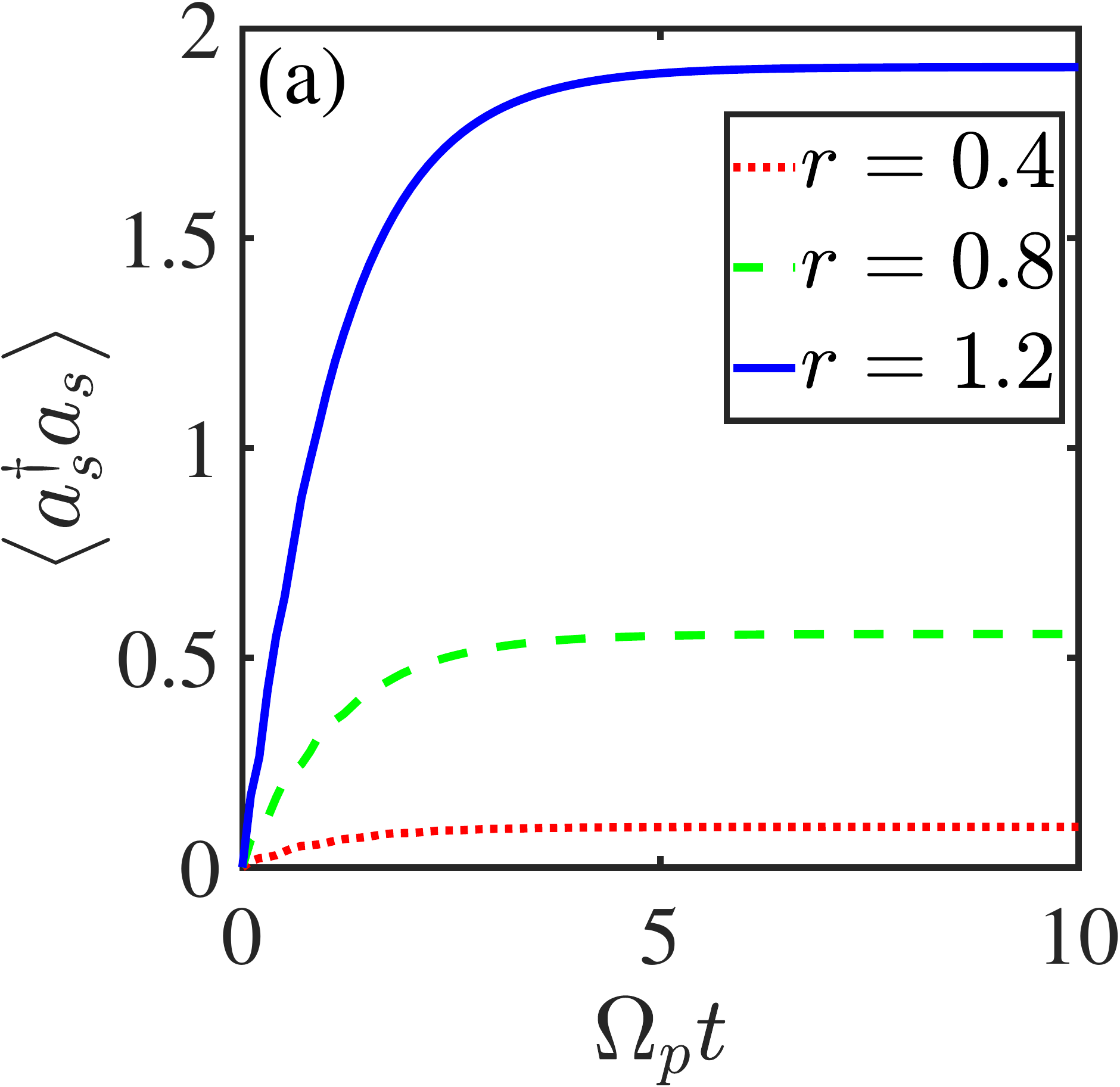}}
	\subfigure{
		\label{Fig4:b}  \includegraphics[width=0.23\textwidth]{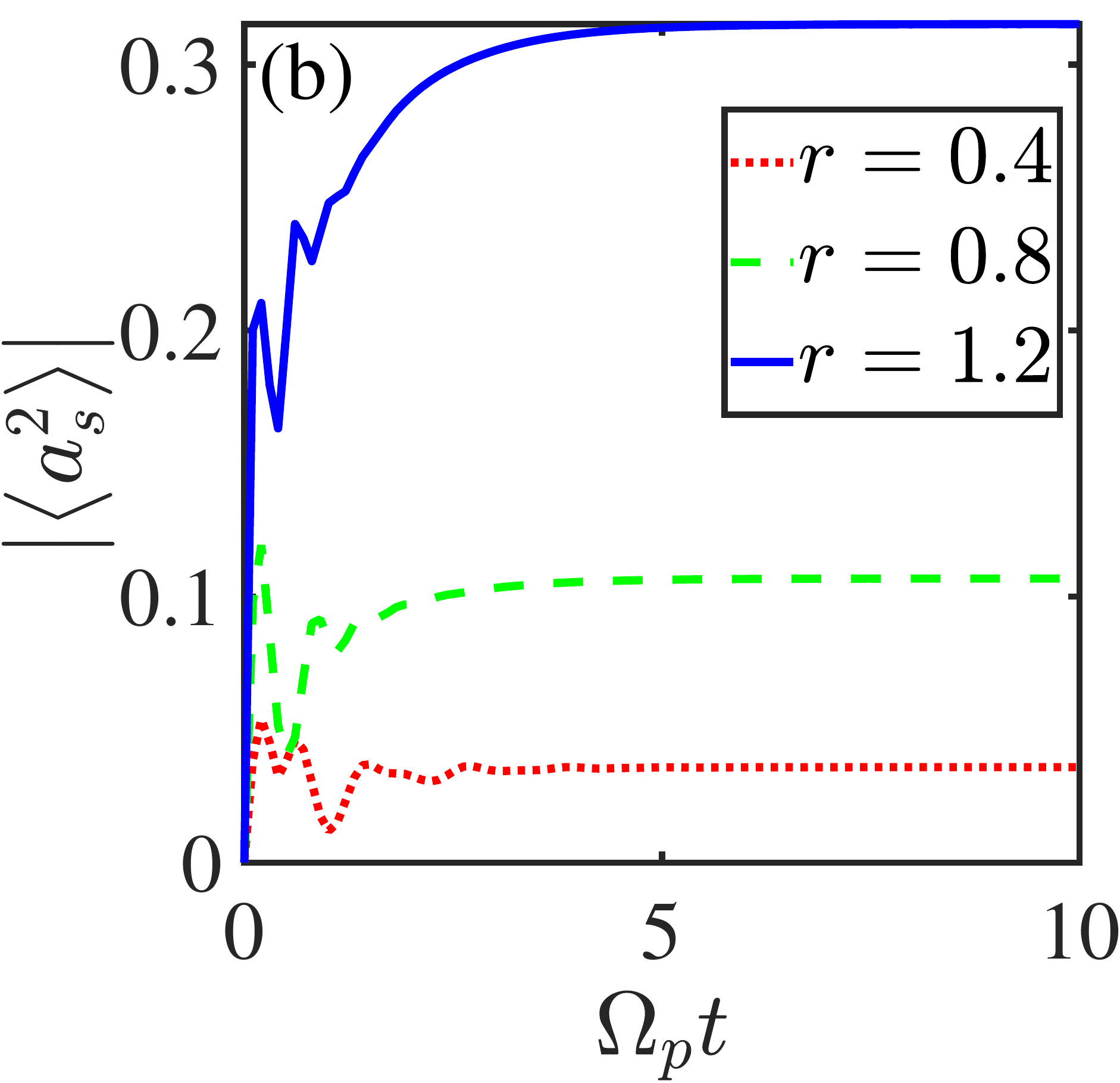}}
	\caption{The time evolutions of (a) the mean photon number $\braket{a_s^{\dagger} a_s}$ and (b) the quantum fluctuations $\left |\braket{a_s^2}\right |$ of the single-atom-cavity QED system for squeezing strength $r=0.4, 0.8, 1.2$, within the squeezed frame. Here, the initial state is assumed to be $\ket{g,0_s}$.~Furthermore, we assume that $g_0=5\kappa$~and$~\gamma/\kappa=1$.}
	\label{Fig4}
\end{figure}

In Fig.$~\ref{Fig4}$, we plot the time evolution of the mean photon number $\braket{a_s^{\dagger} a_s}$ and the quantum fluctuations $ \left |\braket{a_s^2}\right |$ of the single-atom-cavity QED system in the squeezed frame. We find that the mean squeezed-photon number $\braket{a_s^{\dagger} a_s}$ and the quantum fluctuations $\left |\braket{a_s^2}\right |$ vary with time and then gradually approach the stationary values. In the following, the study  will be carried out based on the steady-state~\cite{PhysRevA.42.1737}, and we will denote the steady-state mean by $\braket{o}_{\bar{ss}}$, where the subscript ``$\bar{ss}$" denotes the ``steady-state".

\begin{figure}[!htbp]\centering
	\subfigure{
		\label{Fig5:a}  \includegraphics[width=0.23\textwidth]{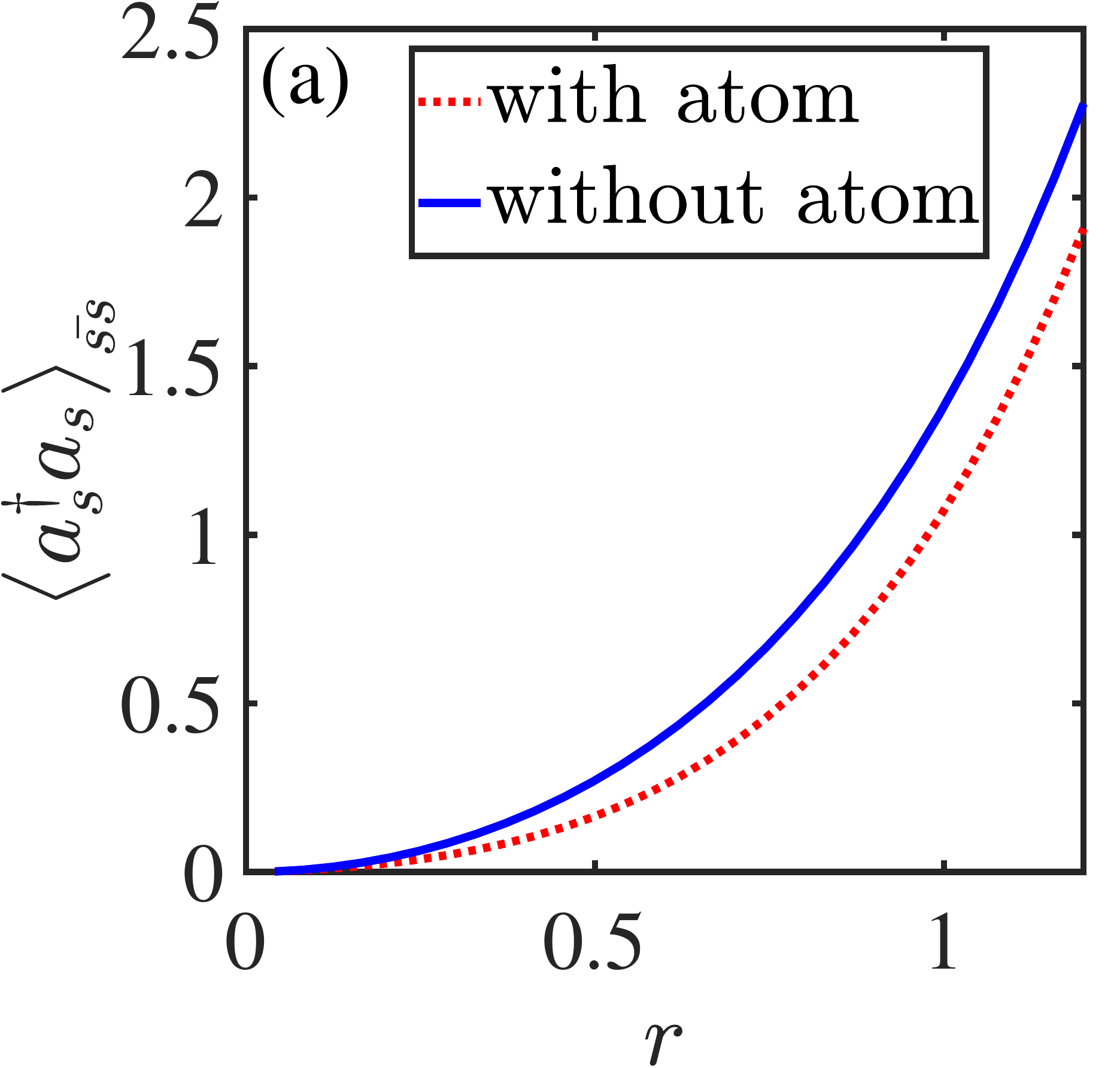}}
	\subfigure{
		\label{Fig5:b}  \includegraphics[width=0.23\textwidth]{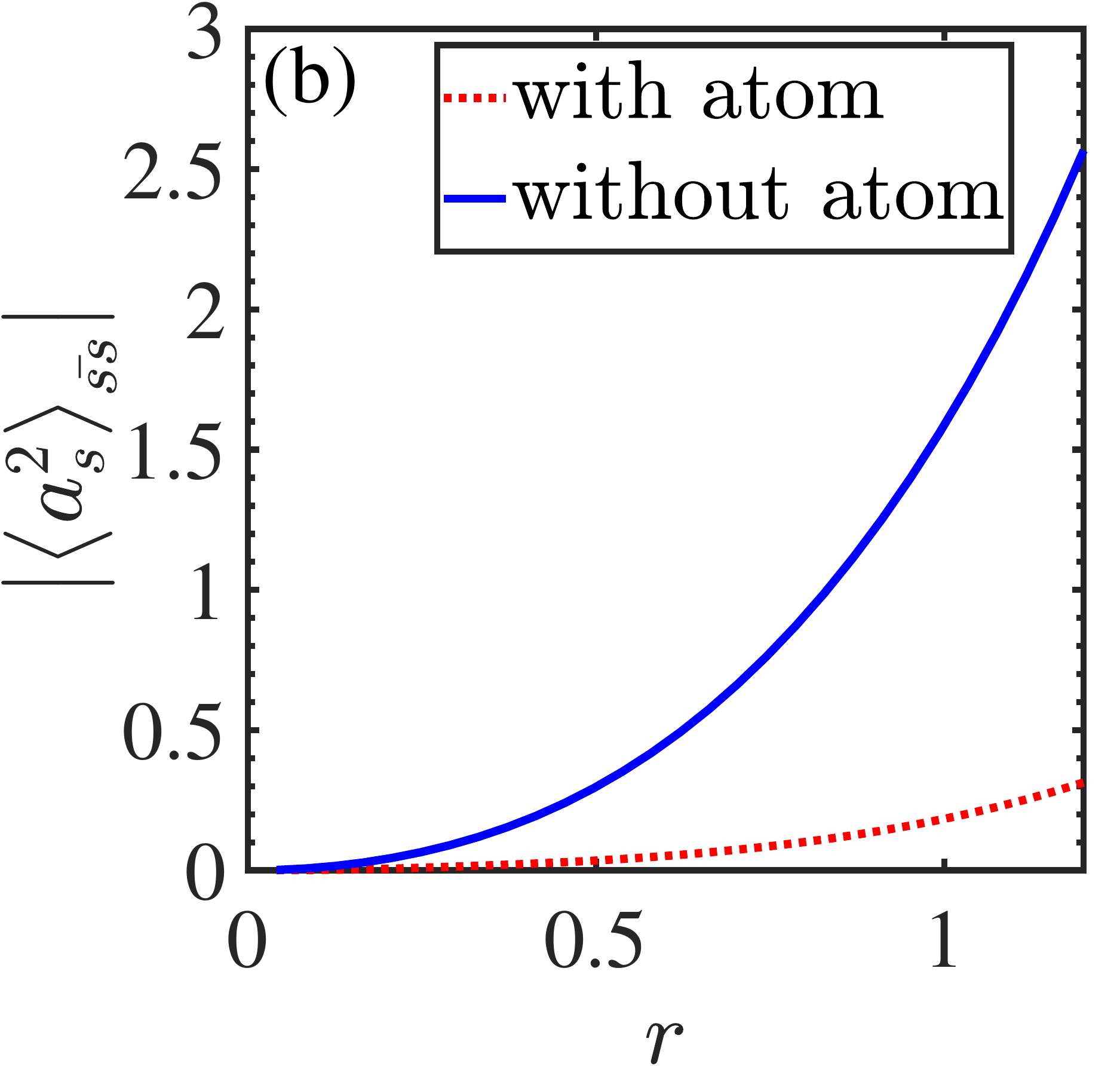}}
	\caption{(a) The mean photon number $\braket{a_s^{\dagger}a_s}_{\bar{ss}}$ versus squeezing strength $r$ in the squeezed frame, calculated from the master equation in Eq.~($\ref{eqn11}$). (b) The quantum fluctuations $\left |\braket{a_s^{2}}_{\bar{ss}}\right |$ versus squeezing strength $r$ in the squeezed frame. Here, the initial state is assumed to be $\ket{g,0_s}$. Furthermore, we assume that $g_0=5\kappa$~and$~\gamma/\kappa=1$.}
	\label{Fig5}
\end{figure}

 In Fig.$~\ref{Fig5}$, we plot the mean photon number $\braket{a_s^{\dagger}a_s}_{\bar{ss}}$ and the quantum fluctuations $\left |\braket{a_s^{2}}_{\bar{ss}}\right |$ as a function of the squeezing parameter $r$, referring to Ref.~\cite{Bao:19}.~We can see from Fig.$~\rm\ref{Fig5:a}$ that, the mean photon number $\braket{a_s^{\dagger}a_s}_{\bar{ss}}$ in the squeezed cavity increases with the increase of the squeezing parameter $r$. The mean photon number $\braket{a_s^{\dagger}a_s}_{\bar{ss}}$ of the single-atom-cavity QED system is going to be less than that of the empty cavity.~This could be explained by the fact that the atom-cavity coupling and atomic dissipation generate additional transition pathways, as shown in Fig.$~\ref{Fig3}$.~For a single-atom-cavity QED system, two photons are produced by squeezed light.~One of them is absorbed for the transition of energy levels, while the other is radiated out of the cavity by the spontaneous emission of the atom. In Fig.$~\rm\ref{Fig5:b}$, we can see that the quantum fluctuations $\left |\braket{a_s^{2}}_{\bar{ss}}\right |$ of the single-atom-cavity QED system are much smaller than that of the empty cavity, because the quantum fluctuations are suppressed by atomic excitations.

\begin{figure}[!htbp]\centering
	\subfigure{
		\label{Fig6:a}  \includegraphics[width=0.23\textwidth]{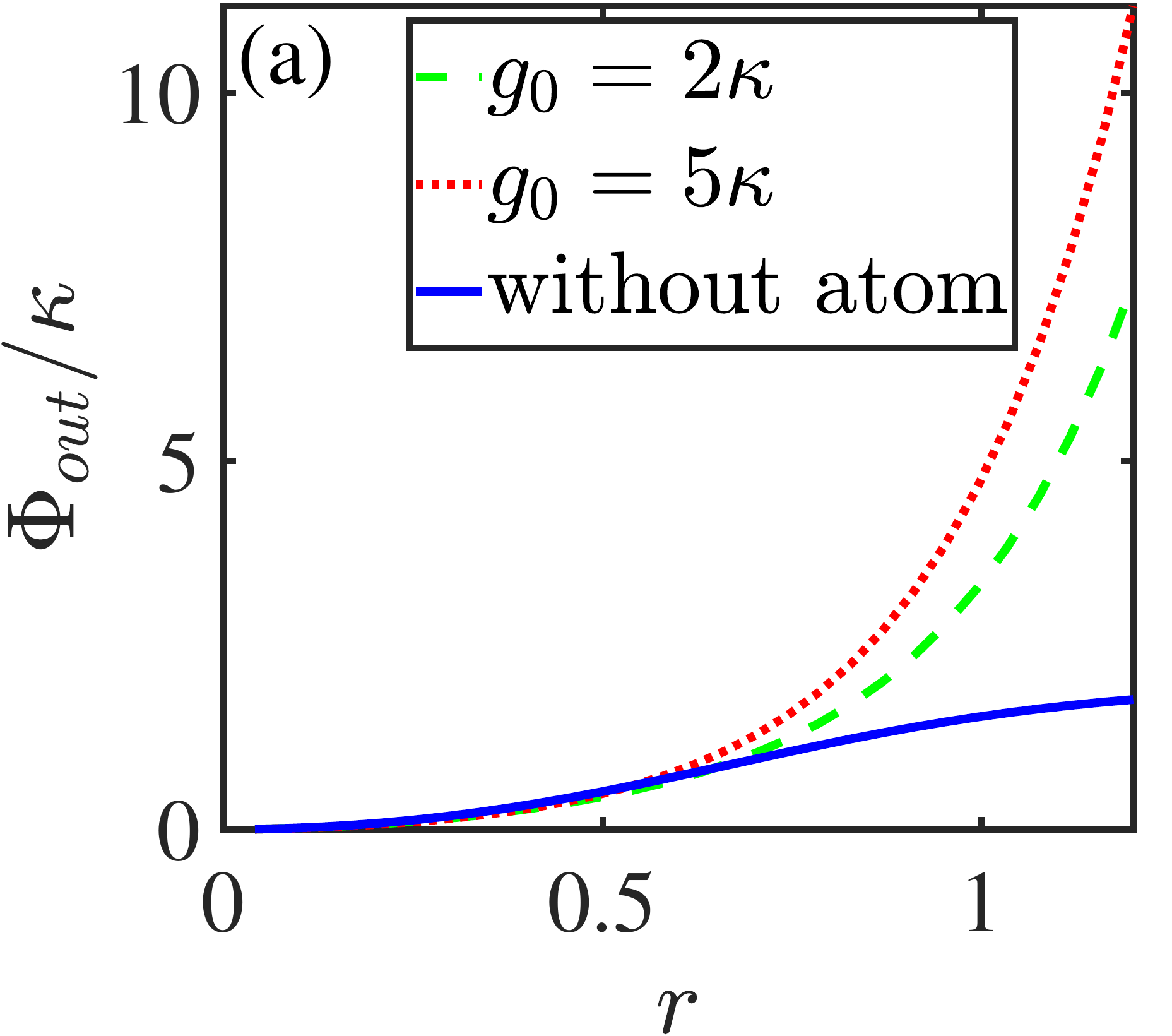}}
	\subfigure{
		\label{Fig6:b}  \includegraphics[width=0.23\textwidth]{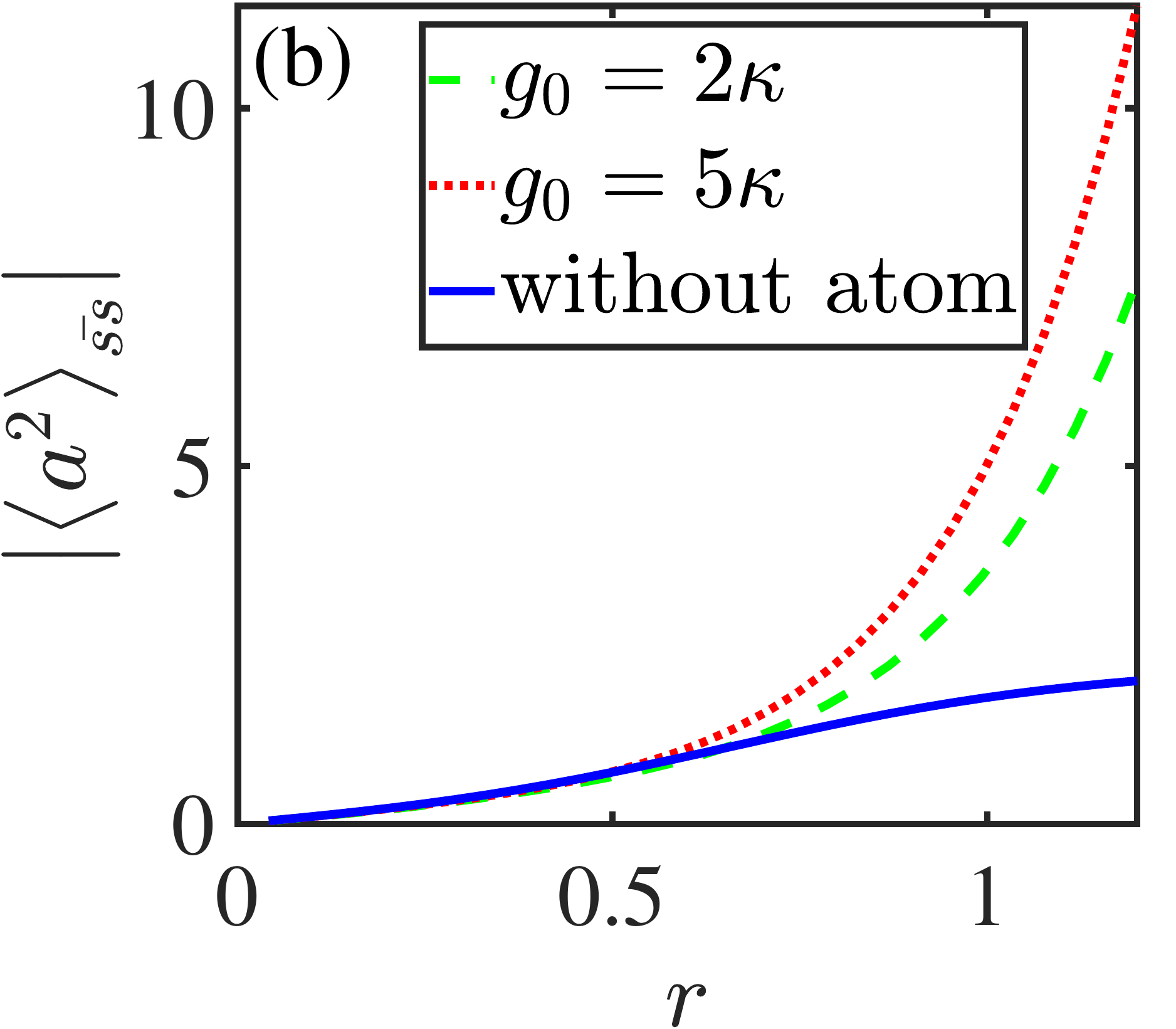}}
	\caption{(a) The output photon flux $\Phi_{\rm out}$ and (b) the quantum fluctuations $\left |\braket{a^2}_{\bar{ss}}\right |$ versus squeezing strength $r$ in the laboratory frame. The blue solid curve represents the case of the empty cavity. The green dashed curve and the red dotted curve denote the cases of the single-atom-cavity QED system with coupling strengths $g_0=2\kappa$ and $g_0=5\kappa$, respectively.}
	\label{Fig6}
\end{figure}

For convenience, following discussions will focus on the laboratory frame.~Having obtained the mean photon number $\braket{a_s^\dagger a_s}_{\bar{ss}}$ and the quantum fluctuations $\left |\braket{a_s^{2}}_{\bar{ss}}\right |$ in the squeezed frame, according to the Bogoliubov transformation, we can calculate
the steady-state intracavity mean photon number $\braket{a^\dagger a}_{\bar{ss}}$ and the quantum fluctuations $\left |\braket{a^{2}}_{\bar{ss}}\right |$ in the laboratory frame,
\begin{equation} \label{eqn14}
	\begin{aligned}
		\braket{a^\dagger a}_{\bar{ss}}&=\sinh^2r+\braket{a_s^\dagger a_s}_{\bar{ss}}\cosh(2r)-{\rm Re}[\braket{a_s^2}_{\bar{ss}}]\sinh(2r),\\
		\left |\braket{a^2}_{\bar{ss}}\right |&=\left |\sinh^2r\braket{a_s^{\dagger 2}}_{\bar{ss}}+\cosh^2r\braket{a_s^2}_{\bar{ss}}\right.\\
		&\left.-\braket{a_s^\dagger a_s}_{\bar{ss}}\sinh(2r)-\sinh r\cosh r\right |.
	\end{aligned}
\end{equation}
According to the input-output relationship~\cite{PhysRevA.78.013640,PhysRevA.100.062501}, the output photon flux is
\begin{equation} \label{e15}
	\Phi_{\rm out}=\kappa\braket{a^\dagger a}_{\bar{ss}}.
\end{equation}
On the basis of Eqs.~($\ref{eqn14}$) and ($\ref{e15}$), we plot the output photon flux $\Phi_{\rm out}$ and the quantum fluctuations $\left |\braket{a^2}_{\bar{ss}}\right |$ versus squeezing strength $r$ in the laboratory frame in Figs.$~\rm\ref{Fig6:a}$ and$~\rm\ref{Fig6:b}$, respectively.~We can see from Figs.$~\rm\ref{Fig6:a}$ and$~\rm\ref{Fig6:b}$ that with the increase of the squeezing strength $r$, the output photon flux $\Phi_{\rm out}$ and the quantum fluctuations $\left |\braket{a^2}_{\bar{ss}}\right |$ of the single-atom-cavity QED system are different from those of the empty cavity. The output photon flux $\Phi_{\rm out}$ and the quantum fluctuations $\left |\braket{a^2}_{\bar{ss}}\right |$ of the empty cavity slightly grow as the squeezing parameter $r$ increases (see the blue solid curve).~While the output photon flux $\Phi_{\rm out}$ and the quantum fluctuations $\left |\braket{a^2}_{\bar{ss}}\right |$ of the single-atom-cavity QED system rapidly increase with $r$ increasing (see the green dashed curve and the red dotted curve). In particular, when $r\textgreater1$, the output photon flux $\Phi_{\rm out}$ and the quantum fluctuations $\left |\braket{a^2}_{\bar{ss}}\right |$ of the single-atom-cavity QED system are significantly different from those of the empty cavity.~The difference between the output photon flux $\Phi_{\rm out}$ (the quantum fluctuations $\left |\braket{a^2}_{\bar{ss}}\right |$) of the single-atom cavity QED system and the output photon flux $\Phi_{\rm out}$ (the quantum fluctuations $\left |\braket{a^2}_{\bar{ss}}\right |$) of the cavity is larger than that in Ref.~\cite{Bao:19}. Therefore, we believe that the present scheme is easier to determine whether or not there is an atom in the cavity than Ref.~\cite{Bao:19}. That is to say, the experimental phenomena of the present scheme are more obvious than those of Ref.~\cite{Bao:19}.~Furthermore, as demonstrated by the green dashed curve and the red dotted curve, the atom-cavity coupling strength has a significant influence on the output photon flux $\Phi_{\rm out}$ and the quantum fluctuations $\left |\braket{a^2}_{\bar{ss}}\right |$. That is, the greater the coupling strength $g_0$ is, the greater the output photon flux $\Phi_{\rm out}$ and the quantum fluctuations $\left |\braket{a^2}_{\bar{ss}}\right |$ are.

In a word, the coupling between the atom and the cavity has significant effects on the output photon flux $\Phi_{\rm out}$ and the quantum fluctuations $\left |\braket{a^2}_{\bar{ss}}\right |$. Moreover, such an effect is significant and can be observed experimentally.~Therefore, such an output photon flux $\Phi_{\rm out}$ and the quantum fluctuations $\left |\braket{a^2}_{\bar{ss}}\right |$ possess valuable applications in detecting whether a single atom is in the cavity or not, which is one of the important results of the scheme.

\begin{figure}[!htbp]\centering
	\subfigure{
		\label{Fig7:a}  \includegraphics[width=0.38\textwidth]{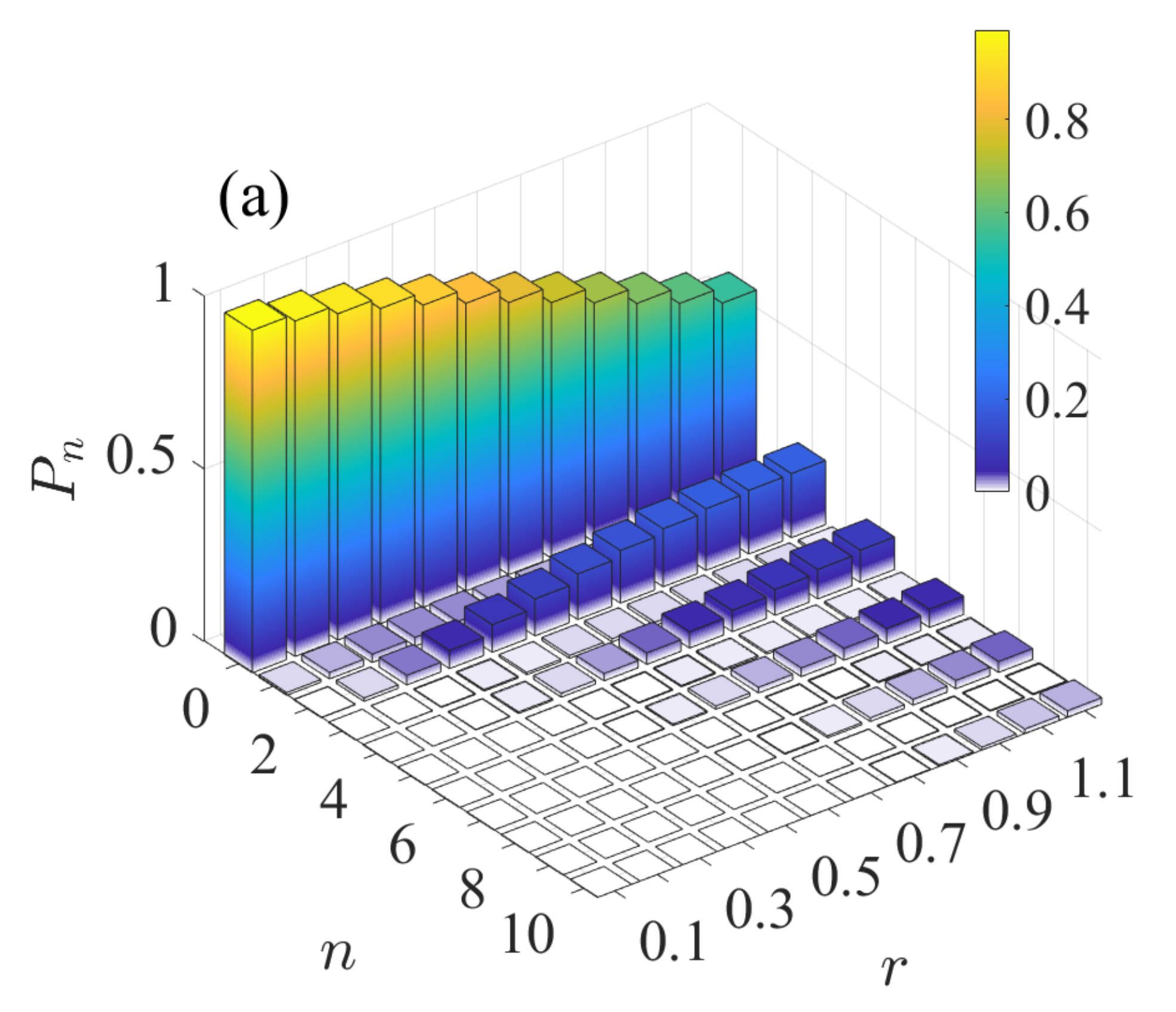}}\\
	\subfigure{
		\label{Fig7:b}  \includegraphics[width=0.38\textwidth]{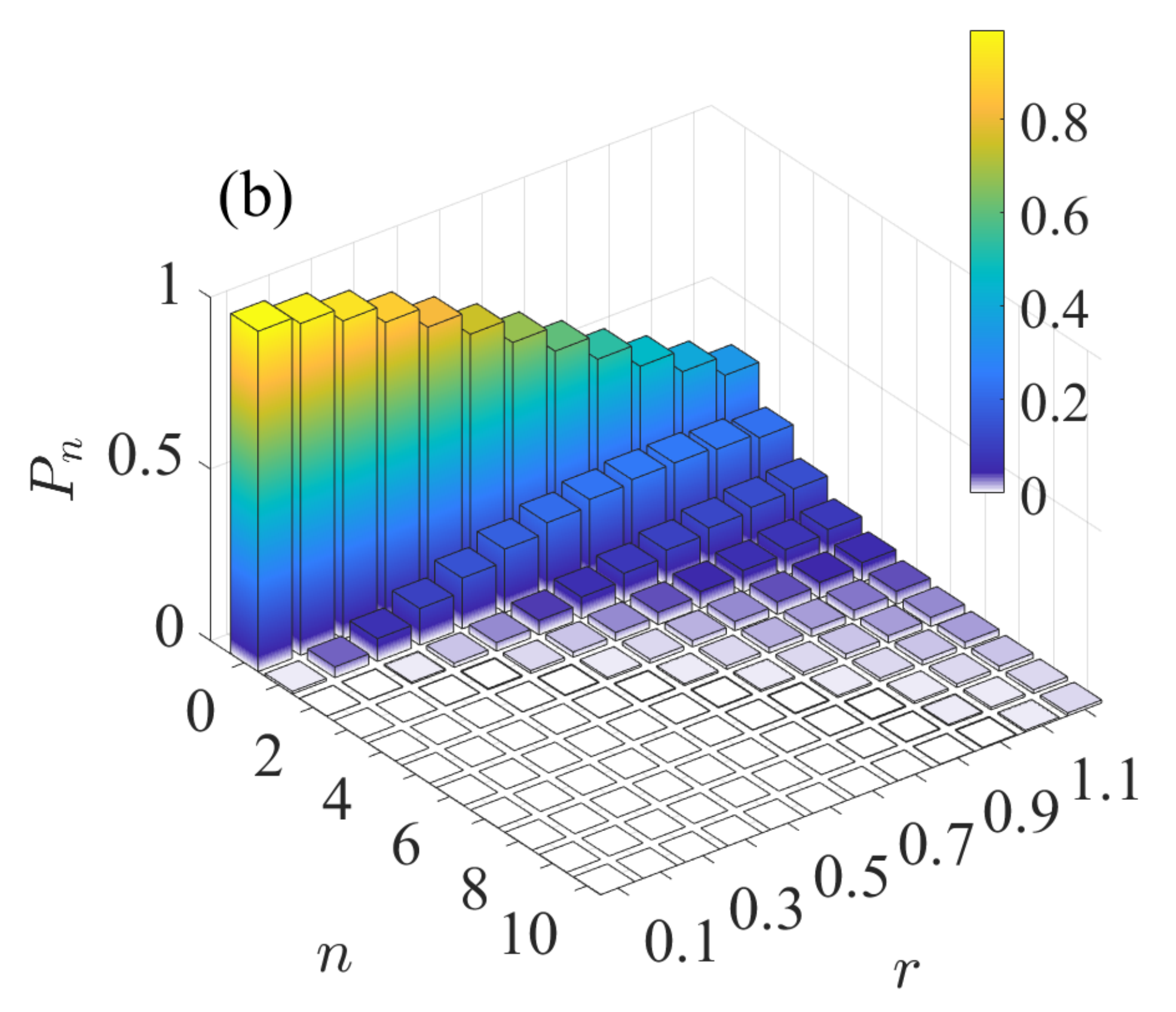}}
	\caption{Photon distribution $P_n$ for different values of $n$ and squeezed strength $r$ in the laboratory frame. Figures (a) and (b) represent the photon distribution probability of the empty cavity and the single-atom-cavity QED system, respectively. Here, the parameters are $g_0=2\kappa$~and$~\kappa=5\gamma$. In addition, $n$ represents the $n$th squeezed Fock state $\hat{S}\ket{n}$.}
	\label{Fig7}
\end{figure}

\begin{figure}[!htbp]\centering
	\scalebox{0.5}{\includegraphics[width=0.96\textwidth]{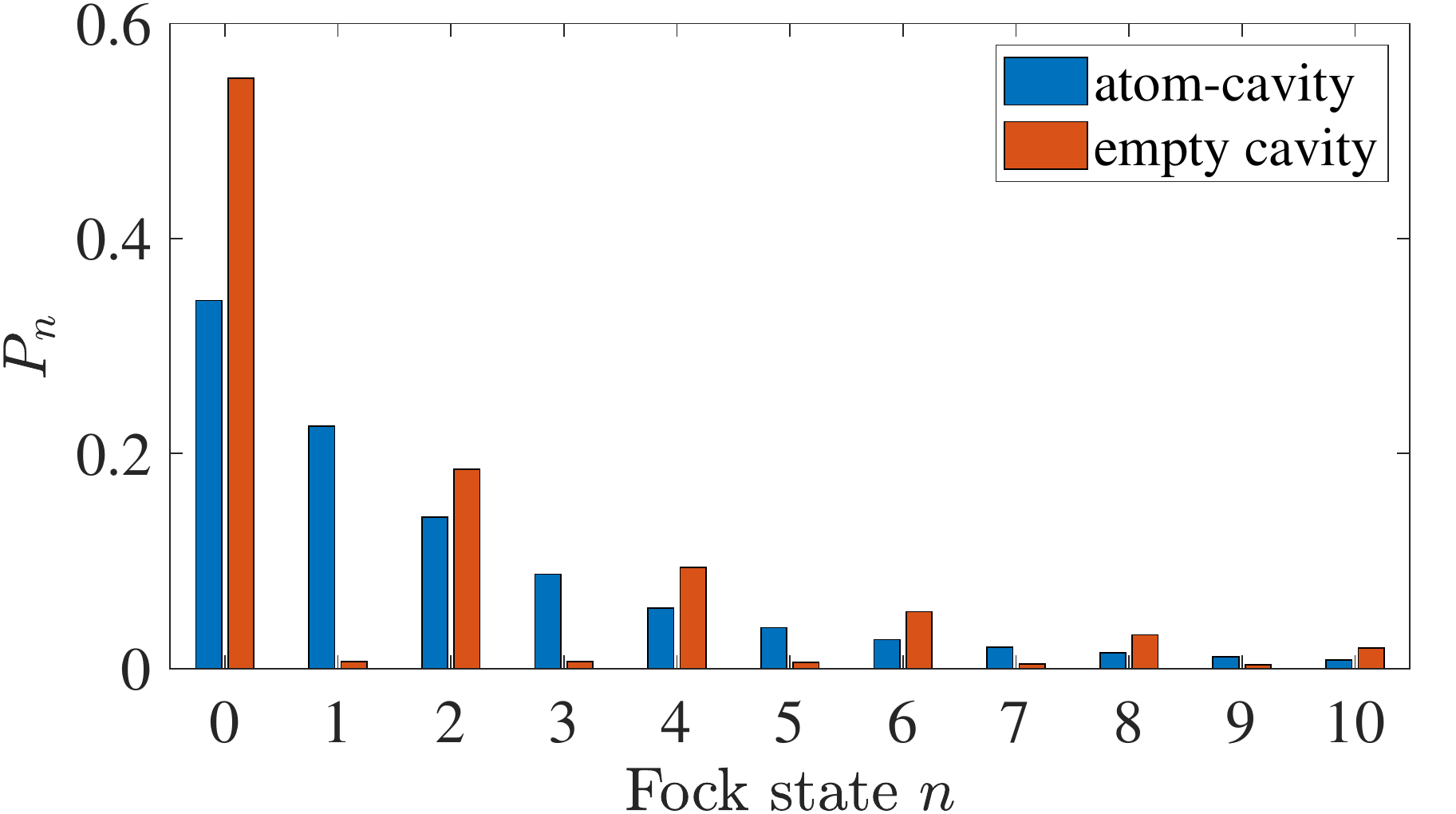}} \caption{Photon distribution in the laboratory frame. The red and blue bars represent the photon distribution of empty cavity and single-atom-cavity QED system, respectively. The parameters are $g_0=2\kappa,~\kappa=5\gamma,$ and $r=1.2$. Here, $n$ represents the $n$th squeezed Fock state $\hat{S}\ket{n}$.}
	\label{Fig8}
\end{figure}

Apart from the output photon flux $\Phi_{\rm out}$ and the quantum fluctuations $\left |\braket{a^2}_{\bar{ss}}\right |$, the photon probability distributions of the empty cavity and the single-atom-cavity QED system are also the object of our study. The photon probability distributions of the empty cavity and the single-atom-cavity QED system for different values of squeezed Fock state $n$ and squeezed strength $r$ are given in Figs.$~\rm\ref{Fig7:a}$ and$~\rm\ref{Fig7:b}$, respectively.~We can see from Fig.$~\rm\ref{Fig7:a}$ that when there is no atom in the cavity, photons are mainly distributed in the even squeezed Fock states. This is because the $\chi^{(2)}$ nonlinear medium driven by an external driving field produces a two-photon effect, resulting in the formation or annihilation of photons in pairs in the cavity.~By comparing Figs.$~\rm\ref{Fig7:a}$ and$~\rm\ref{Fig7:b}$, one can find that the photon distribution of the odd squeezed Fock states increases obviously when the atom is trapped in the cavity, which is caused by the cavity-atom coupling and the spontaneous emission of the atom. As shown in Fig.$~\ref{Fig8}$, when the squeezed strength $r$ is constant, the difference between the photon probability distribution of the empty cavity and that of the single-atom-cavity QED system is especially obvious. In other words, the photon probability distribution of the empty cavity shows a trend of oscillation with the increase of $n$, while the photon probability distribution of the single-atom-cavity QED system shows a trend of slow decrease with the increase of $n$.~Therefore, the photon probability distribution can be used as one of the criteria to determine whether an atom is successfully trapped inside the cavity or not. Note that the sum of photon distribution probability is bound by 1. Because the highly excited states $|n>10\rangle_{s}$ of the squeezed cavity mode are mostly unexcited, we have ignored them in Figs. $~\ref{Fig7}$ and $~\ref{Fig8}$.

\begin{figure}[!htbp]\centering
	\subfigure{
		\label{Fig9:a}  \includegraphics[width=0.23\textwidth]{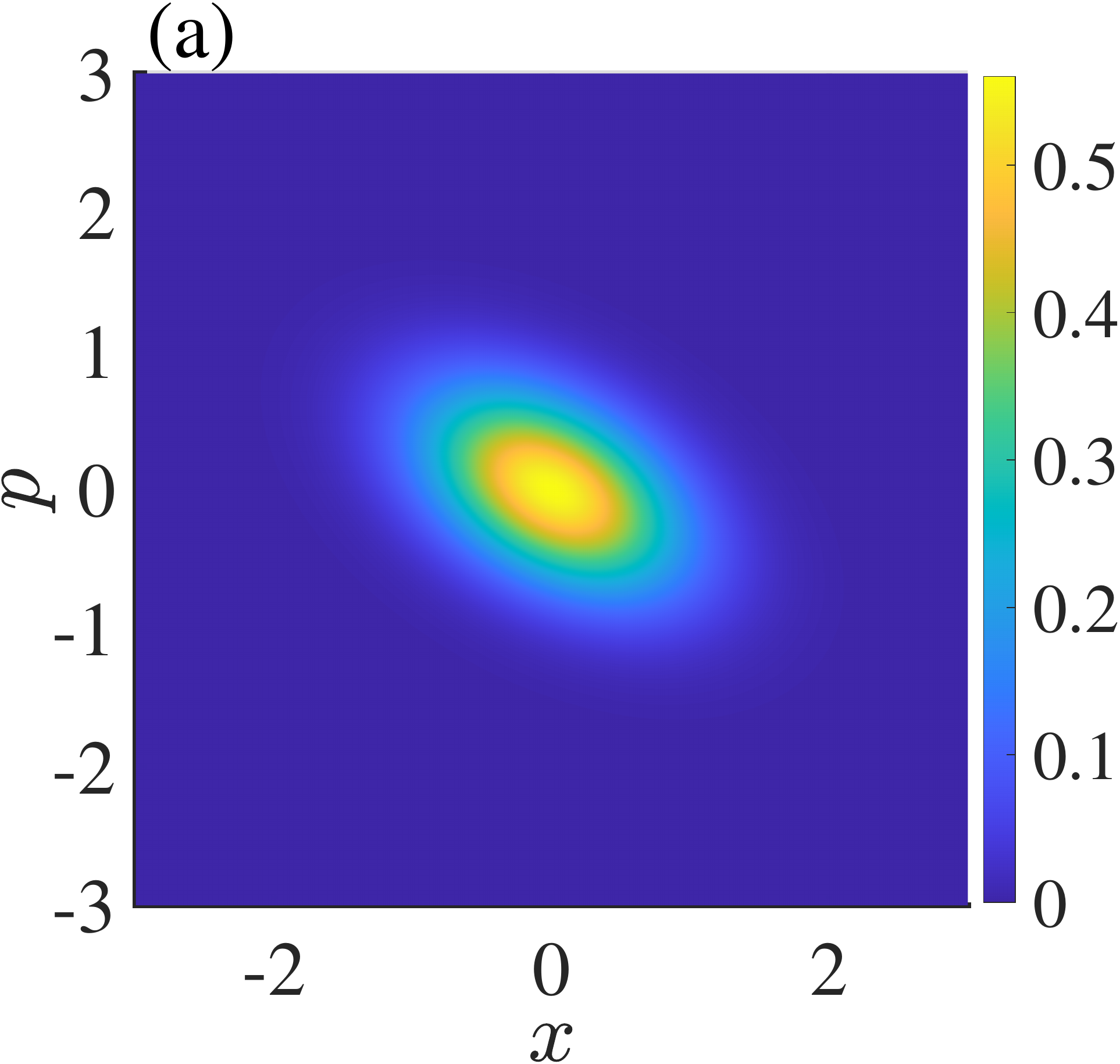}}
	\subfigure{
		\label{Fig9:b}  \includegraphics[width=0.23\textwidth]{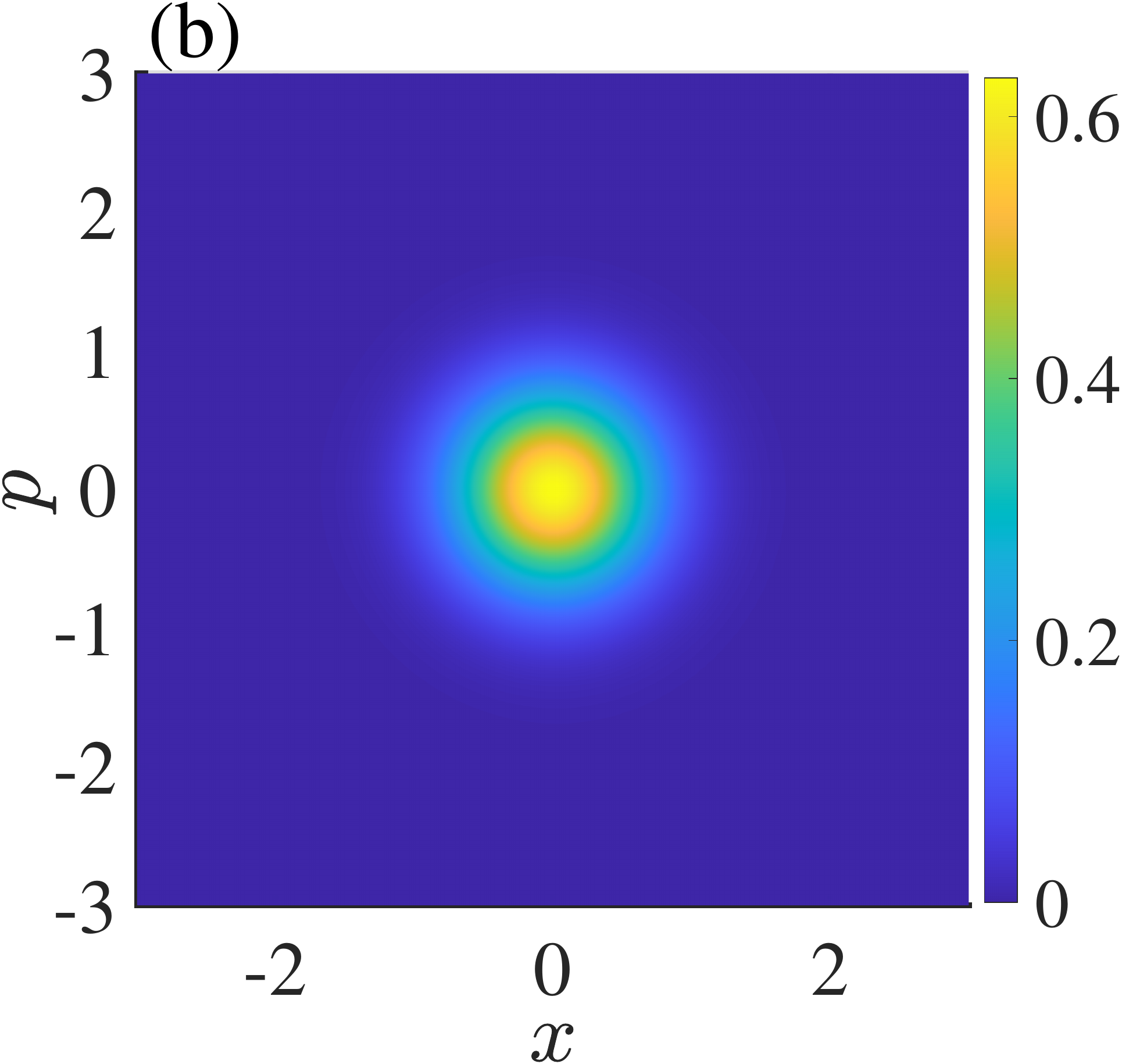}}
	\caption{Figures (a) and (b) show the Wigner function of the cavity fields in the laboratory frame of the empty cavity and the single-atom-cavity QED system, respectively. Here, we assume that $g_0=5\kappa,~r=1,$ and $\gamma/\kappa=1$. $x$ and $p$ represent the real and imaginary parts of the coherent state, respectively.}
	\label{Fig9}
\end{figure}

The coupling between atom and cavity also has a significant effect on the quantum statistical properties of the system.~We use the Wigner function $W(\alpha)=(2/\pi)Tr[D(\alpha)(-1)^{a^{\dagger}a}D^{\dagger}(\alpha)\rho]$~\cite{PhysRev.40.749} of the cavity field to characterize the quantum statistical properties of the system~\cite{Bao:19}.~In the laboratory frame, the Wigner function is calculated from the solution
of Eq.~($\ref{eqn2}$). Figures.$~\rm\ref{Fig9:a}$ and $\rm\ref{Fig9:b}$ show the Wigner function of the empty cavity and the single-atom-cavity QED system, respectively. As shown in Fig.$~\rm\ref{Fig9:a}$, for the empty cavity, when the $\chi^{(2)}$ nonlinear medium is driven by the additional driving field, the two-photon effect comes into being and acts on the state to create some sort of squeezed states~\cite{gerry2005introductory}. The Wigner function of the empty cavity is suppressed in one direction, which means that the cavity mode is squeezed. Furthermore, the Wigner function of the single-atom-cavity QED system is shown in Fig.$~\rm\ref{Fig9:b}$. This can be interpreted as the excitation and spontaneous emission of atoms preventing the generation of a squeezed field. Compared with the case of the empty cavity, it is clear that the presence of an atom significantly changes the Wigner function of system in the laboratory frame. As a consequence, the Wigner function of the cavity can be used as a signal to determine the existence of atoms.

\section{CONCLUSION}\label{CONCLUSION}
In this paper, a protocol for detecting a single atom in a cavity using the $\chi^{\left( 2\right) }$ nonlinear medium is proposed. Because the $\chi^{(2)}$ nonlinear medium is driven by an external laser field, the cavity mode will be squeezed, and the coupling between the atom and the squeezed cavity mode is exponentially enhanced. Compared with the cases of the empty cavity, the output photon flux, quantum fluctuations, quantum statistical properties, and photon number distribution of the single-atom-cavity QED system are significantly affected by the atom and the $\chi^{(2)}$ nonlinear medium.~Moreover, the greater the coupling strength between the atom and the cavity, the more obvious the effect is.~These allow to determinately sense an atom in a cavity.~The proposed protocol possesses many advantages, such as controllable squeezing strength and squeezed-cavity-mode frequency, and exponential enhancement of atom-cavity coupling strength.~In addition, using nonlinear medium to produce squeezed light has been realized experimentally~\cite{Ast:13,Serikawa:16,PhysRevLett.117.110801,SCHNABEL20171}. Experimentally, a parametric gain of $10\log_{10}\left[\exp(2r)\right] \sim 20~\rm dB$ (corresponding to $r \sim 2.3$) has been achieved~\cite{doi:10.1126/science.aaw2884}, and $\sim 30$ dB has also been predicted under experimentally feasible conditions~\cite{Clark2017,PhysRevLett.121.173601,Murch2013}.~Our protocol is a supplement to the existing single atom detection protocols in cavities, and we hope it can be promissing for atomic detection in other quantum systems.

\begin{acknowledgments}
This work was supported by the National
Natural Science Foundation of China under Grant Nos. 11575045,
11874114, and 11674060, the Natural Science Funds for Distinguished
Young Scholar of Fujian Province under Grant 2020J06011
and Project from Fuzhou University under Grant JG202001-2. Y.-H.C. is supported by the Japan Society for the Promotion of Science (JSPS) KAKENHI Grant No.~JP19F19028.
\end{acknowledgments}
\appendix
\section{DERIVATION OF SQUEEZING PARAMETER}\label{A}
The detailed derivation of squeezing parameter Eq.~(\ref{e3}) is as follows.~After substituting the Bogoliubov squeezing transformation Eq.~(\ref{e5}) into the nonlinear Hamiltonian $H_{\rm NL}$ in Eq.~(\ref{eqn1}) for degenerate parametric amplification, the nonlinear Hamiltonian $H_{\rm NLS}$ in squeezed frame becomes
\begin{equation} \label{A1}
	\begin{aligned}
		H_{\rm NLS}&=\Delta_{c}[\cosh^2(r)a_s^{\dagger}a_s-\sinh(r)\cosh(r)e^{-i\theta_p}a_s^{\dagger 2}\\
		&-\sinh(r)\cosh(r)e^{i\theta_p}a_s^{2}+\sinh^2(r)a_s a_s^{\dagger} ]\\	
		&+\frac{1}{2}\Omega_p[\cosh^2(r)e^{i\theta_p}a_s^{2}-\sinh(r)\cosh(r)a_sa_s^{\dagger}\\
		&-\sinh(r)\cosh(r)a_s^{\dagger}a_s+\sinh^2(r)e^{-i\theta_p}a_s^{\dagger 2}+\rm H.c.].
	\end{aligned}
\end{equation}
According to commutation relation $[a_s,a_s^\dagger]=1$, we can get
\begin{equation} \label{A2}
	\begin{aligned}
		H_{\rm NLS}&=\Delta_{c}[\cosh^2(r)a_s^{\dagger}a_s-\sinh(r)\cosh(r)e^{-i\theta_p}a_s^{\dagger 2}\\
		&-\sinh(r)\cosh(r)e^{i\theta_p}a_s^{2}+\sinh^2(r)a_s^{\dagger}a_s + \sinh^2(r)]\\	
		&+\frac{1}{2}\Omega_p[\cosh^2(r)e^{i\theta_p}a_s^{2}-\sinh(r)\cosh(r)a_s^{\dagger}a_s\\
		&-\sinh(r)\cosh(r)-\sinh(r)\cosh(r)a_s^{\dagger}a_s\\
		&+\sinh^2(r)e^{-i\theta_p}a_s^{\dagger 2}+\rm H.c.].
	\end{aligned}
\end{equation}
To diagonalize $H_{\rm NLS}$ , the following conditions need to be met
\begin{equation} \label{A3}
	\begin{aligned}
		&\Delta_{c}[\cosh^2(r)+\sinh^2(r)]-2\Omega_p\sinh(r)\cosh(r)=\omega_s,\\
		&\frac{1}{2}\Omega_p[\sinh^2(r)+\cosh^2(r)]-\Delta_{c}\sinh(r)\cosh(r)=0.\\
	\end{aligned}
\end{equation}
So by using the double angle formula, we can simplify Eq.~(\ref{A3}) to get
\begin{equation} \label{A4}
	\begin{aligned}
		\Delta_{c}\cosh(2r)-\Omega_p\sinh(2r)=\omega_s,
	\end{aligned}
\end{equation}
\begin{equation} \label{A5}
	\begin{aligned}
		\Delta_{c}\sinh(2r)+\Omega_p\cosh(2r)=0.
	\end{aligned}
\end{equation}
After substituting the exponential form of hyperbolic function into Eqs.~(\ref{A4}) and (\ref{A5}), we can obtain 
\begin{equation} \label{A6}
	\begin{aligned}
		\Delta_{c}(e^{2r}+e^{-2r})-\Omega_p(e^{2r}-e^{-2r})=\omega_s,
	\end{aligned}
\end{equation}
\begin{equation} \label{A7}
	\begin{aligned}
		\Delta_{c}(e^{2r}-e^{-2r})-\Omega_p(e^{2r}+e^{-2r})=0.
	\end{aligned}
\end{equation}
From Eqs.~(\ref{A6}) and~(\ref{A7}) we can derive
\begin{equation} \label{A8}
	\begin{aligned}
		r=\frac{1} {4}\ln\left(\frac{\Omega_p+\Delta_{c}} {\Omega_p-\Delta_{c}}\right),
	\end{aligned}
\end{equation}
\begin{equation} \label{A9}
	\begin{aligned}
		\omega_s=\sqrt{\Delta_c^2-\Omega_p^2}.
	\end{aligned}
\end{equation}

\bibliographystyle{apsrev4-1}
\bibliography{ref}
\end{document}